\documentclass[12pt]{iopart}

\usepackage{microtype}
\usepackage{graphicx}
\usepackage[english]{babel}
\usepackage{url}
\begin{document}

\author{D.~E.~Kahana\dag\  and S.~H.~Kahana\ddag}
\address{\dag 31 Pembrook Drive, Stony Brook, NY 11790, USA}
\address{\ddag Physics Department, Brookhaven National Laboratory\\
   Upton, NY 11973, USA}
\eads{\mailto{kahana@bnl.gov}, \mailto{dek@bnl.gov}}

\title{LHCb $P_c^+$ Resonances as Molecular States.}
\date{\today}

\begin{abstract}

Recent experiments at the LHCb detector have once again raised the possibility
that quark bags may exist containing more than three quarks.  Specifically the
LHCb collaboration points to evidence for hadronic resonances decaying into
J/$\psi$ and proton: the $P_c(4450)^+$ and $P_c(4380)^+$. Here we put forth a
case that a reasonable description of these states is possible as molecular
resonances of a $p$ and a J/$\psi$.  Our model seems to accommodate the
observed states and their measured widths, both the lower lying, broader,
negative parity state and the higher lying, narrow positive parity state. If
these resonances do indeed exist one might envision a rich spectroscopy of
such pentaquark states waiting to be discovered, though this would deepen the
mystery of their absence in earlier hadron spectroscopy. Perhaps the presence
of the heavy charmed quarks, as well as the lighter u,d and s quarks, is the
determining factor.

\end{abstract}

\vspace{2pc}
\vspace{1pc}


\maketitle

\section{Introduction} 

We proposed previously~\cite{kahana}, a description of the experimental
observations~\cite{ThetaPlus1,ThetaPlus2} that were made about a decade ago of
a possible $\Theta^+$ pentaquark, as a molecule formed of $K^+$ and $n$ in a
high relative angular momentum resonant state, as the most likely description
for the narrow experimentally observed structure pointing to the existence of
the state. However, we warned that verification of the data pointing to the
existence of the $\Theta^+$ was essential and indeed the evidence did not hold
up under further scrutiny~\cite{Hicks}. The present experiment at
LHCb~\cite{LHCb} suggesting the observation of 5-quark structure appears to be
of much higher quality and hence we now re-examine the possibility of
describing the current observations in the same manner, modeling the $P_c^+$
as a molecular resonance having {\it atomic} components that are $p$ and
J/$\psi$.

The recent data taken at LHCb~\cite{LHCb} demonstrating the, albeit rare,
production of a five quark resonance and its decay into proton and J/$\psi$
motivates a theoretical search for a possible molecular or shape
resonance. The large mass of one of the decay products favours the
construction of such a picture using a non-relativistic potential
model~\cite{kahana}. This approach is quite successful, and easily can produce
two states in reasonable agreement with the observed excitation energies,
widths and parities suggested by the experimenters. In our description, the
states possess orbital angular momentum $l=3,4$ with the higher angular
momentum state being considerably narrower and higher in excitation
energy. This of course would also accounts for the change in parity between
the two observed states. The narrowness of the higher state is assured mostly
by the higher centrifugal barrier that exists for the larger angular momentum
state.

The experimental data, Figs(2a,2b), in the LHCb submission~\cite{LHCb},
provide strong evidence for a narrow feature and apparently equal statistical
justification for a lower mass but considerably broader resonance.  The two
states seem inextricably linked in the molecular framework.

In the dynamical approach of Hasenfratz and Kuti~\cite{hasenfratz}, referred
to in our earlier discussion of pentaquarks as molecular states~\cite{kahana},
the surface of each component particle is deformable and susceptible to
surface oscillations~\cite{bohr} which can be expanded in spherical harmonics,
each labeled by an orbital angular momentum $l$.

The Hasenfratz-Kuti model employs a surface potential~\cite{bohr} with
strength proportional to

\begin{equation}
V_S(l)= -(l-1)(l+2)\rho_0^2\sigma, 
\end{equation}

\noindent where $\rho_0$ is an average bag radius and $\sigma$ the surface
tension. Such a coupling might arise from a comparable dependence in the
overall effective J/$\psi$-$p$ interaction via particle-vibrational
coupling. One outcome of our search is an $l$-dependence for the $l=3$ and
$l=4$ potential depths whose ratio, $V_S(3)/V_S(4)\sim 1/2$, is close to that
anticipated in Eq(1), given equal surface tensions    and radii.

We explored a variety of molecular potentials, both volume and surface forms,
employing the code GAMOW ~\cite{Gamow} to find the resonant states. For the
volume potential we tried a Woods-Saxon and an explicit
surface forms. The strong centripetal potential for these relatively high orbital
angular momentum states, however, renders both volume and surface forms
effectively surface-peaked; so the choice of attractive potential
is not actually that important.

We also explored the possibility that an $l=2$ resonance exists and found a
possible feature in the range of excitation energies $E\sim 100-200$ MeV with
variable width, perhaps be less easily observable.
Even ignoring the putative explicit $l$ dependence of the
potential above, $l=0$ and $l=1$ resonances are absent a consequence of the
lower centrifugal barriers.

Purely bag-like or soliton-like pentaquark models face a
common theoretical problem. Their lowest lying states will have low relative
angular momentum between the constituent quarks and so will be connected to
low angular momentum $l=0,1$ outgoing waves~\cite{capstick} thus
acquiring very large widths.

\section{Results and Comments}

Since the various choices of potential give essentially equivalent results we
present results only for a particular form of surface potential. These are
displayed in Figs(1,2). The relative parities are experimentally opposite in
sign and are reproduced in our modeling by the differing orbital angular
momentum states $l=3$ and $l=4$ with correspondingly assigned total angular
momenta $J=3/2$ and $J=5/2$, and appropriate $L-S$ coupling. For simplicity,
spin dependent forces were omitted in our treatment.

To create a surface potential permitting differing interior and exterior
diffusivities we choose a surface potential in the form:
\begin{equation}
V_S(l)=V_s \,\,\, x^{\alpha-1}(1-x)^{\beta-1},
\end{equation}
a simple finite range potential formed from a product of two power laws, with
exponents $\alpha-1$, and $\beta-1$. The radial scale of the potential is
given by the parameter $r_0$, with $x=r/r_0$.  For a small inner diffusivity
and an even steeper exterior rise we take $\alpha=9$, $\beta=5$ and for the
scale parameter $r_0=0.85$ fm.

The depth of the well is adjusted in the code GAMOW to obtain resonances at
the known and observed excitation energies of the resonant states. The
potential well depths and the widths of the states are then predictions of the
model. Fixing the $l=4$ state excitation energy at $E=410$ MeV corresponding
to an experimental total mass $M = 4450$ MeV we find, Fig(1), a width $\Gamma
\simeq 50$ MeV and a maximum depth $V_S \simeq 2360$ MeV. Correspondingly we
obtain a width of $\Gamma \simeq 142$ MeV and depth $V_S=1410$ MeV for the
appreciably broader $l=3$ resonance at $E=340$ MeV excitation energy, as
illustrated in Fig(2). The ratio of these well depths is then close to that
which follows from Eq(1). The wave functions of the resonant states, both the
real and imaginary parts thereof, are also presented in these figures.

The wave functions for the two orbital states allow the atomic molecular
components, J/$\psi$ and $p$ respectively some $0.3$ and $0.8$ fm in radius,
to exist comfortably as independent particles in the molecular state. One can
nevertheless expect some distortion of the two bag surfaces. The outer radius
of a resonant state is of course arbitrary, since the inside wavefunctions are
matched by GAMOW smoothly onto outgoing spherical waves to solve the resonant
condition on the scattering matrix. Clearly the effective inner region extends
to more than some $1.5$ fm. Averaging the potential over this limited region
of its effectiveness leads to a considerably reduced mean well depth, and the
dynamics can be safely described within a non-relativistic framework. This is
evident in the smooth behaviour of the wavefunctions seen in both Fig(1) and
Fig(2), indicating a rather small kinetic energy is present in either the
$l=3$ or the $l=4$ state.

The production of the quarks involved in the initial $P_c^+$, in Fig(1b) of
the experimental publication~\cite{LHCb}, is a short-time, hard QCD process,
while the formation of the final observed hadrons involves much longer times
$\tau \sim 1$ fm/c, {\it i.e.} it is a soft process. Of course, as indicated
by the LHCb Collaboration, in Fig(1a) and Fig(1b) the formation of the $P_c^+$
likely takes place in two stages, first through the production of a five quark
$\Lambda_b^0$ state followed by the signature splitting, after the generation
of a $u\bar{u}$ pair from the vacuum, into final observed hadrons $K^-$ and
the accompanying ``5-quark'' state $P_c^+$. The latter subsequently decays
into the identifying J/$\psi$ and $p$ hadrons. Perturbative QCD may be able to
describe the production of the initial quarks but calculating wave functions
for the final states is clearly, at present, possible only in a
phenomenological model such as that employed here and like that used successfully 
to describe the J/$\psi$ itself~\cite{gottfried}.

Given the much longer formation time of the molecular state, which necessarily
requires the involvement of low momentum transfer processes, the initial quark
production details aside from, of course, the number and type of quarks
present should have little effect on the eventual molecular state that
results.

As to related 5-quark resonances that one might expect exist, given the LHCb
observations, certainly states with $b\bar{b}$ replacing $c\bar{c}$ in the
J/$\psi$ could be searched for; or, for that matter, states obtained by
replacing the J/$\psi$ by a higher bound charmonium state, for example the
$\psi'$. As we previously indicated, it is also possible that an $l=2$
resonance could be present in the existing experiment. But this would be seen
at substantially lower excitation energy and therefore would be distinct from
the $l=3$ and $l=4$ states. It remains a puzzle why more than 3-quark hadrons
are not evident at lower masses in the large pre-existing literature on hadron
spectroscopy; this may, as pointed out above, be related to the presence in
the $P_C^+$ of more massive charm or, possibly, if such additional states are
found, bottom quarks.
\clearpage

\begin{figure}[p]
\includegraphics*[trim=-2.0cm 0 0 0]{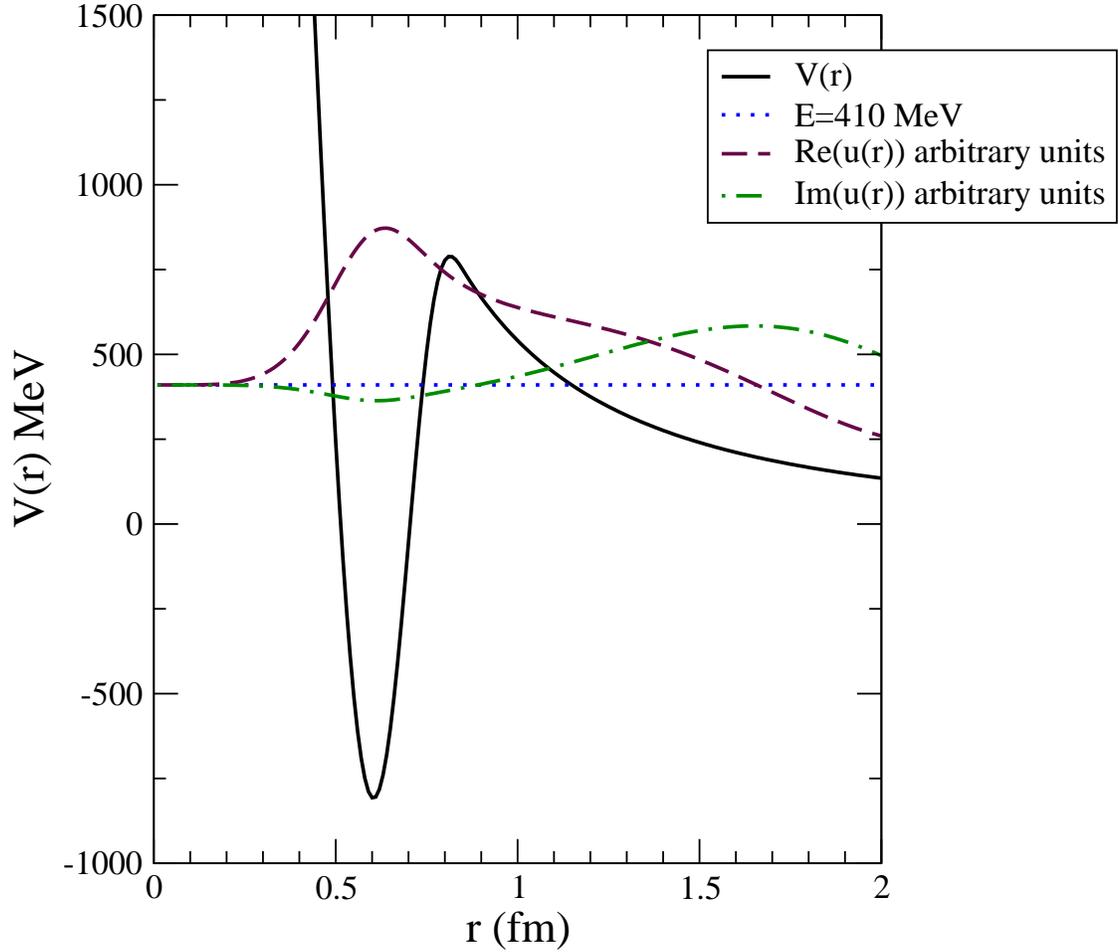}
\caption[]{$l=4$ Resonance: Surface Potential and Wave Function. In our scheme
  this positive parity state corresponds to the narrow feature seen in the
  data from LHCb.  We display the summed centripetal and quark-surface
  vibrational terms. The maximum depth of the inter-hadronic potential is
  searched on to yield the desired excitation energy. The resultant maximum
  depth $V_s \simeq 2360$ MeV and width $\Gamma \simeq 50$ MeV are then
  results of this search. The predicted width $\Gamma \simeq 50$ MeV for the
  $l=4$ state is consistent with the experimental value.  Also shown are the
  real and imaginary wave functions yielded in GAMOW~\cite{Gamow}. Their
  smooth variation inside a radius of say $1.5$--$2.0$ fm confirms the
  validity of the non-relativistic calculation. The real wave function which
  effectively extends uniformly to at least $1.5$ fm indicates the atomic
  components of the two hadrons fit comfortably inside the proposed molecule.}
\label{fig:Fig.(1)}
\end{figure}
\clearpage

\begin{figure}[p]
\includegraphics*[trim=-2.0cm 0 0 0]{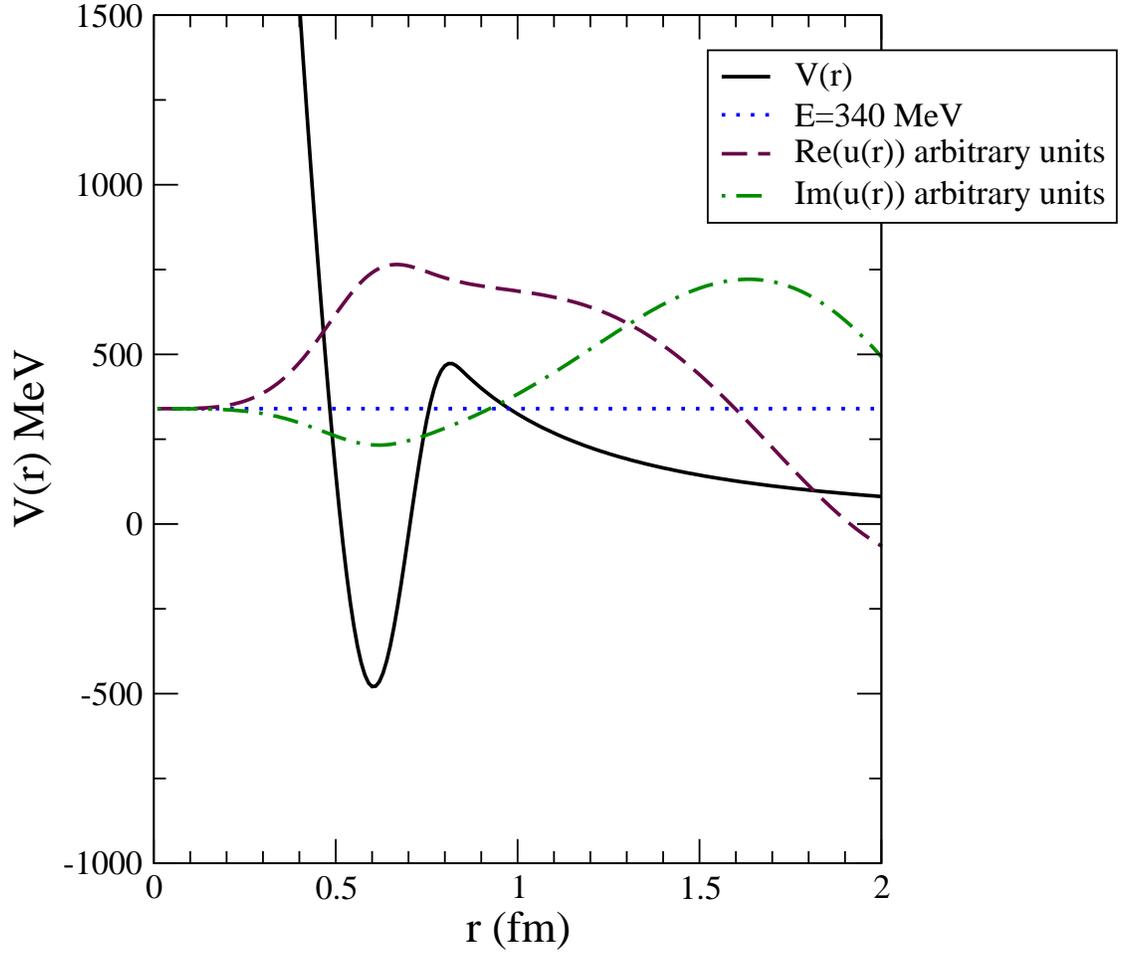}
\caption[]{A similar plot for the 340 MeV $l=3$ resonance, clearly the
  negative parity partner of the state in Fig(1). The derived width for this
  state $\Gamma \simeq 142$ MeV is in reasonable agreement with that of this
  much broader state observed in the LHCb data. Also, the maximum depth
  of the surface potential derived, $V_s=1410$ MeV, is close in ratio to that
  obtained for the $l=4$ state predicted by the $l$-dependence in Eq(1), {\it
    i.e.} $\sim 1/2$.}
\label{fig:Fig.(2)}
\end{figure}
\clearpage

\ack This manuscript has been authored under the US DOE grant
NO. DE-AC02-98CH10886.

\section*{References}
\bibliography{pentaquarkc}

\end{document}